\documentclass[10pt]{article}
\usepackage{latexsym,graphicx}
\usepackage{latexsym,graphicx,epsf}
\newcommand{\be}{\begin{equation}}
\newcommand{\ee}{\end{equation}}
\def\n{\noindent}
\catcode `\@=11 \catcode `\@=12
\begin{document}
\begin{center}
\large{\bf {FRW UNIVERSE WITH VARIABLE $G$ and $\Lambda$-TERMS }} \\
\vspace{10mm}
\normalsize{ANIRUDH PRADHAN\footnote{Corresponding author}, AJAY KUMAR SINGH} \\
\vspace{5mm} \normalsize{Department of 
Mathematics, Hindu Post-graduate College, Zamania-232 331, Ghazipur, India} \\
\normalsize{$^1$E-mail: pradhan@iucaa.ernet.in, acpradhan@yahoo.com}\\
\vspace{5mm}
\normalsize{SAEED OTAROD} \\
\vspace{5mm} \normalsize{Department of Physics, Yasouj University, Yasouj, Iran} \\
\normalsize{sotarod@mail.yu.ac.ir, sotarod@yahoo.com}\\
\vspace{5mm}
\end{center}
\vspace{10mm}
%\date{}
%\maketitle
\begin{abstract}
A new class of exact solutions of Einstein's field equations with
a perfect fluid source, variable gravitational coupling $G$ and 
cosmological term $\Lambda$ for FRW spacetime is obtained by considering 
variable deceleration parameter models for the universe. The nature of 
the variables $G(t)$, $\Lambda(t)$ and the energy density $\rho(t)$ 
have been examined for three cases : (i) exponential (ii) polynomial and 
(iii) sinusoidal form. The special types of models for dust, Zel'dovich and 
radiating universe are also mentioned in all these cases. The behaviour of 
these models of the universe are also discussed in the light of recent 
supernovae Ia observations.
\end{abstract}
\smallskip
\n Key words : Cosmology; Deceleration parameter; Variable gravitational and 
cosmological terms.\\\
\n PACS: 98.80.Jk, 98.80.Es\\
%\newpage
%%%%%%%%%%%%%%%%%%%%%%%%%%%%%%%%%%%%%%%%%%%%%%%%%%%%%%%%%%%%%%%%%%%%%%%%%%%%%%%%%%%%%
%%%%%%%%%%%%%%%%%%%%%%%%%%%%%%%%% section 1 Introduction %%%%%%%%%%%%%%%%%%%%%%%%%%%
\section{Introduction}
In Einstein's General Relativity (GR), there are two parameters, the 
gravitational constant $G$ and the cosmological constant $\Lambda$ which 
are considered to be time-independent \cite{ref1}. A possible time variable 
$G$ was suggested by Dirac \cite{ref2} and this has been extensively discussed in 
literature \cite{ref3,ref4}. Since its introduction, its significance has 
been studied from time to time by various workers \cite{ref5} $-$ \cite{ref7}. 
In modern cosmological theories, the cosmological constant remains a focal 
point of interest. A wide range of observations now compellingly suggest 
that the universe possesses a non-zero cosmological constant \cite{ref8}. 
In the context of quantum field theory, a cosmological term corresponds to the 
energy density of vacuum. The birth of the universe has been attributed to an 
excited vacuum fluctuation triggering off an inflationary expansion followed 
by the super-cooling. The release of locked up vacuum energy results in 
subsequent reheating. The cosmological term, which is measure 
of the energy of empty space, provides a repulsive force opposing the 
gravitational pull between the galaxies. If the cosmological term exits, 
the energy it represents counts as mass because  mass and energy are equivalent. 
If the cosmological term is large enough, its energy plus the matter in the 
universe could lead to inflation. Unlike standard inflation, a universe with 
a cosmological term would expand faster with time because of the push from 
the cosmological term \cite{ref9}. Some of the recent discussions on the 
cosmological constant ``problem'' and on cosmology with a time-varying
cosmological constant by Ratra and Peebles \cite{ref10}, Dolgov
\cite{ref11}$-$\cite{ref13} and Sahni and Starobinsky \cite{ref14} point out 
that in the absence of any interaction with matter or radiation, the cosmological 
constant remains a ``constant''. However, in the presence of interactions with 
matter or radiation, a solution of Einstein equations and the assumed equation 
of covariant conservation of stress-energy with a time-varying $\Lambda$ can be 
found. This entails that energy has to be conserved by a decrease in the energy 
density of the vacuum component followed by a corresponding increase in the energy 
density of matter or radiation [see also Weinberg \cite{ref15} and Carroll, Press 
and Turner \cite{ref16}]. \\

Recent observations by Perlmutter {\it et al.} \cite{ref17} and 
Riess {\it et al.} \cite{ref18} strongly favour a significant and a positive 
value of $\Lambda$ with magnitude $\Lambda(G\hbar/c^{3}) \approx 10^{-123}$. 
Their study is based on more than $50$ type Ia supernovae with 
red-shifts in the range $0.10 \leq z \leq 0.83$ and these suggest Friedmann 
models with negative pressure matter such as a cosmological constant 
$(\Lambda)$, domain walls or cosmic strings (Vilenkin \cite{ref19}, Garnavich 
{\it et al.} \cite{ref20}). Recently, Carmeli and Kuzmenko \cite{ref21} have shown 
that the cosmological relativistic theory predicts the value for cosmological 
constant $\Lambda = 1.934\times 10^{-35} s^{-2}$. This value of ``$\Lambda$'' 
is in excellent agreement with the recent estimates of the High-Z Supernova Team 
and Supernova Cosmological Project (Garnavich {\it et al.} \cite{ref20}; 
Perlmutter {\it et al.} \cite{ref17}; Riess {\it et al.} \cite{ref18}; 
Schmidt {\it  et al.} \cite{ref22}). These observations suggest on accelerating  
expansion of the universe. \\

Since Dirac \cite{ref2} first considered the possibility of a variable 
$G$, there have been numerous modifications of general relativity to 
make $G$ vary with time \cite{ref23}. Recently a modification linking 
the variation of $G$ with that of $\Lambda$ has been considered within 
the framework of general relativity by a number of workers \cite{ref24}
$-$\cite{ref29}. \\

The Einstein's field equations are a coupled system of non-linear differential 
equations. We seek physical solutions to the field equations for their 
applications of relevance to cosmology and astrophysics. In order to solve the 
field equations we normally assume a form for the matter content and killing 
vector symmetries by space time \cite{ref30}. Solutions to the field equations 
may also be generated by applying a law of variation for Hubble's parameter 
proposed by Berman \cite{ref31}. In simple cases the Hubble law yields a constant 
value of deceleration parameter. It is worth observing that most of the well-known 
models of Einstein's theory and Brans-Deke theory with curvature parameter $k = 0$, 
including inflationary models have a constant deceleration parameter as borne out 
by the studies of several authors \cite{ref31}$-$\cite{ref36}. But redshift magnitude 
test has had a chequered history. During the 1960s and the 1970s, it was used to draw 
very categorical conclusions. The deceleration parameter $q_{0}$ was then claimed to 
lie between $0$ and $1$ leading to the suggestion that the universe is decelerating. 
But today, the view is entirely different. Observations (Knop et al. \cite{ref37};
Riess et al., \cite{ref38}) of Type Ia Supernovae (SNe) allow us to probe the
expansion history of the universe leading to the conclusion that the expansion of 
the universe is accelerating. So we can consider cosmological models with
variable cosmological term and deceleration parameter. Vishwakarma and Narlikar 
\cite{ref39} and Virey et al. \cite{ref40} have reviewed the determination of the 
deceleration parameter from Supernovae data. \\

Recently Pradhan and Otarod \cite{ref41} have studied a model of the universe 
with time dependent deceleration parameter and $\Lambda$-term in 
presence of perfect fluid. This study has prompted us to  focus upon the problem 
of establishing a formalism for understanding  the relativistic evolution for 
a time dependent deceleration parameter in an expanding universe with variable $G$ and 
$\Lambda$ in presence of perfect fluid. This paper is organized as follows. 
We present in Section 2 the metric and the field equations. In Section $3$ we deal 
with a general solution. In Sections $4$, $5$ and $6$ we deal with the solutions in 
exponential, polynomial and sinusoidal form. Finally in Section $7$ we mention our 
main conclusions. 
%%%%%%%%%%%%%%%%%%%%%%%%%%%%%%%%%%%%%%%%%%%%%%%%%%%%%%%%%%%%%%%%%%%%%
%%%%%%%%%%%%%%%%%%%%%%%%%%%%%%%  SECTION 2  %%%%%%%%%%%%%%%%%%%%%%%%%%

\section{The metric and field  equations}
In standard coordinates $(x^{a}) = (t, r, \theta, \phi)$ the Robertson-
Walker line element has the form 
\begin{equation}
\label{eq1} ds^{2} = - dt^{2} + S^{2}(t)\left[\frac{dr^{2}}{1 - kr^{2}} + 
r^{2}(d\theta^{2} + \sin^{2}\theta d\phi^{2})\right], 
\end{equation}
where $S(t)$ is the cosmic scale factor. Without loss of generality the 
constant $k$ takes only three values: $0, 1$ or $- 1$. The constant $k$ 
is related to spatial geometry of a 3-dimensional manifold generated by $t$ = 
constant. The Robertson-Walker spacetimes are the standard cosmological models 
supporting observations. For the case of variable cosmological term $\Lambda(t)$ 
and gravitational constant $G(t)$ the Einstein field equation 
\begin{equation}
\label{eq2} G_{ab} + \Lambda g_{ab} = 8\pi G(t) T_{ab}
\end{equation}
yields
\begin{equation}
\label{eq3} \frac{3}{S^{2}}(\dot{S}^{2} + k) = 8\pi G \rho + \Lambda, 
\end{equation} 
\begin{equation}
\label{eq4} 2\frac{\ddot{S}}{S} + \frac{(\dot{S}^{2} + k)}{S^{2}} = 
- 8\pi G p + \Lambda, 
\end{equation}
for the line element (\ref{eq1}) where a dot indicate partial differentiation 
with respect to $t$. From Eqs. (\ref{eq3}) and (\ref{eq4}) we obtain the 
generalized continuity equation 
\begin{equation}
\label{eq5} \dot{\rho} + 3\frac{\dot{S}}{S}\left(\rho + p\right) + \frac{\dot{G}}{G} 
\mu + \frac{\dot{\Lambda}}{8\pi G} = 0.
\end{equation}
This reduces to the conventional continuity equation when $\Lambda$ and $G$ are 
constants. In an attempt to obtain solutions to the field equations we assume, as 
is often done, that the classical conservation law, $T^{ab}_;{b} = 0$, also holds. 
Then Eq. (\ref{eq5}) leads to  
\begin{equation}
\label{eq6} \dot{\rho} + 3\frac{\dot{S}}{S}\left(\rho + p\right), 
\end{equation}
\begin{equation}
\label{eq7} 8\pi \rho \dot{G} + \dot{\Lambda} = 0,
\end{equation}
which facilitate the solution of the field equations. The result (\ref{eq6}) is 
just the conventional continuity equation, and (\ref{eq7}) simply relates $G$ and 
$\Lambda$ and does not explicitly contain the scale factor $S(t)$.
%%%%%%%%%%%%%%%%%%%%%%%%%%%%%%%%%%%%%%%%%%%%%%%%%%%%%%%%%%%%%%%%%%%%%%%%%
%%%%%%%%%%%%%%%%%%%%%%%%%%  SECTION 3  %%%%%%%%%%%%%%%%%%%%%%%%%%%%%%%%%%

\section{Solution of the Field Equations}
Equations (\ref{eq3}) - (\ref{eq4}) are two independent equations in five 
unknowns $S$, $p$, $\rho$, $G$, and $\Lambda$. In this paper we first assume an 
equation of state
\begin{equation}
\label{eq8} p = \gamma \rho, 0 \leq \gamma \leq 1.
\end{equation}
Secondly, on the basis of supernovae searches, we consider the deceleration 
parameter to be variable and set
\begin{equation}
\label{eq9}
q = - \frac{S \ddot S}{{\dot S}^2} = - \left(\frac{\dot H + H^2}{H^2}\right) 
= b \mbox{(variable)},
\end{equation}
where $H = \dot S/S$ is the Hubble parameter. The above equation may be 
rewritten as
\begin{equation}
\label{eq10} \frac{\ddot{S}}{S}+ b \frac{{\dot{S}}^2}{S^{2}} = 0.
\end{equation}
The general solution of Eq. (\ref{eq10}) is given by
\begin{equation}
\label{eq11} \int{e^{\int{\frac{b}{S}dS}}}dS = t + m,
\end{equation}
where $m$ is an integrating constant.

In order to solve the problem completely, we have to choose
$\int{\frac{b}{S}dS}$ in such a manner that Eq. (\ref{eq11})
be integrable.

Without loss of generality, we consider
\begin{equation}
\label{eq12} \int{\frac{b}{S}dS} = {\rm ln} ~ {L(S)},
\end{equation}
and obtain from Eqs. (\ref{eq11}) and (\ref{eq12}) 
\begin{equation}
\label{eq13} \int{L(S)dS} = t + m.
\end{equation}
Of course the choice of $L(S)$, in Eq. (\ref{eq13}), is quite
arbitrary but, since we are looking for physically viable models
of the universe consistent with observations, we consider the
following three cases:
%%%%%%%%%%%%%%%%%%%%%%%%%%%%%%%%%%%%%%%%%%%%%%%%%%%%%%%%%%%%%%%%%%%%%
%%%%%%%%%%%%%%%%%%%%%%%%%%%%%%%  SECTION 4  %%%%%%%%%%%%%%%%%%%%%%%
\section{Solution in the Exponential Form}
Let us consider $ L(S) = \frac{1}{k_{1}S}$, where $k_{1}$ is arbitrary constant.

In this case on integration of Eq. (\ref{eq13}) gives the exact solution 
\begin{equation}
\label{eq14} S(t) = k_{2}e^{k_{1}t}, 
\end{equation}
where $k_{2}$ is an arbitrary constant. \\
Using Eq. (\ref{eq7}) in the differentiated form of (\ref{eq3}), we obtain
\begin{equation}
\label{eq15} 8\pi G \dot{\rho} = - \frac{6k k_{1}}{S^{2}}. 
\end{equation}
From Eqs. (\ref{eq3}), (\ref{eq4}) and (\ref{eq8}), we have
\begin{equation}
\label{eq16} 8 \pi (1 + \gamma) G \rho = \frac{2k}{S^{2}}.   
\end{equation}  
From Eqs. (\ref{eq15}) and (\ref{eq16}), we obtain
\begin{equation}
\label{eq17} \frac{\dot{\rho}}{\rho} = - 3k_{1}(1 + \gamma),   
\end{equation}
which on integration gives 
\begin{equation}
\label{eq18} \rho = A e^{ - 3k_{1}(1 + \gamma)t},  
\end{equation}
where $A$ is an integrating constant. Using Eqs. (\ref{eq14}) and (\ref{eq18}) 
in (\ref{eq16}) we get the value of $G$
\begin{equation}
\label{eq19} G = \frac{ke^{(1 + 3\gamma)k_{1} t}}{4\pi(1 + \gamma)Ak_{2}^{2}}. 
\end{equation}
From Eq. (\ref{eq3}) or (\ref{eq4}), we find $\Lambda$ as
\begin{equation}
\label{eq20} \Lambda = \frac{(1 + 3\gamma)ke^{-2k_{1}t}}{(1 + \gamma)k_{2}^{2}} 
+ 3 k_{1}^{2}. 
\end{equation}
From Eq. (\ref{eq14}), since scale factor can not be negative, we find $S(t)$ 
is positive if $k_{2} > 0$. So in the early stages of the 
universe, i.e., near $t = 0$, the scale factor of the universe had been 
approximately constant and had increased very slowly. Some time later, the 
universe had exploded suddenly and expanded to a large scale. This picture is 
consistent with Big Bang scenario.\\

From Eq. (\ref{eq18}), we note that $\rho(t)$ is a decreasing function of 
time and $\rho > 0$ for all times. From Eq. (\ref{eq19}), we observe that $G$ 
is an increasing function of time. When the universe is required to have expanded 
from a finite minimum volume, the critical density assumption and conservation of 
energy-momentum tensor dictate that $G$ increases in a perpetually expanding 
universe \cite{ref24}. In most variable $G$ cosmologies \cite{ref42,ref43} $G$ is 
a decreasing function of time. But the possibility of an increasing $G$ has also 
been suggested by several authors \cite{ref44}.\\

The behaviour of the universe in this model will be determined by the cosmological 
term $\Lambda$ ; this term has the same effect as a uniform mass density $\rho_{eff} 
= - \Lambda / 4\pi G$, which is constant in space and time. A positive value of 
$\Lambda$ corresponds to a negative effective mass density (repulsion). Hence, we 
expect that in the universe with a positive value of $\Lambda$, the expansion will 
tend to accelerate; whereas in the universe with negative value of $\Lambda$, 
the expansion will slow down, stop and reverse. From Eq. \cite{ref20}, we see 
that the cosmological term $\Lambda$ is a decreasing function of time and it 
approaches a small positive value as time increase more and more. Recent cosmological 
observations (Garnavich et al. \cite{ref19}, Perlmutter et al. \cite{ref16}, 
Riess et al. \cite{ref17}, Schmidt et al. \cite{ref21}) suggest the existence of 
a positive cosmological constant $\Lambda$ with the magnitude 
$\Lambda(G\hbar/c^{3}\approx 10^{-123}$. These observations on magnitude and 
red-shift of type Ia supernova suggest that our universe may be an accelerating 
one with induced cosmological density through the cosmological $\Lambda$-term. 
Thus our model is consistent with the results of recent observations. In 
the following, we consider the variation of $\rho(t)$, $\Lambda(t)$ and $G(t)$ 
with time for three different special cases.\\

The expressions for $\rho(t)$, $\Lambda(t)$ and $G(t)$ corresponding to
$\gamma = 0, 1, 1/3$ are given in {\bf Table 1}. 

\vspace{0.5cm}

\begin{tabular}{|c|c|c|c|} \hline
$\gamma$        &   $\rho(t)$           & $\Lambda(t)$                                   & $G(t)$  \\ \hline
$0$             &   $Ae^{-3k_{1}t} $   & $\frac{ke^{-2k_{1}t}}{k_{2}^2} + 3k_{1}^{2}$   & $\frac{ke^{k_{1}t}}{4\pi A k_{2}^{2}} $        \\  \hline
$ 1$            &  $Ae^{-6k_{1}t} $    & $\frac{2ke^{-2k_{1}t}}{k_{2}^2} + 3k_{1}^{2}$  & $\frac{ke^{4k_{1}t}}{8\pi A k_{2}^{2}} $         \\  \hline    

$\frac{1}{3}$   &  $Ae^{-4k_{1}t} $    & $\frac{3ke^{-2k_{1}t}}{2k_{2}^2} + 3k_{1}^{2} $  &  $\frac{3ke^{k_{1}t}}{16\pi A k_{2}^{2}} $   \\  \hline 

\end{tabular} 
\vspace{0.5cm}

{\bf Table $1$}: Values of $\rho(t)$, $\Lambda(t)$ and $G(t)$ for dust, Zel'dovice 
and radiation exponential models. \\ 

From Table $1$, we observe that in all three models $\rho > 0$ for all times and is 
a decreasing function of time. The cosmological term $\Lambda$ is also decreasing 
function of time and it approaches to a small positive value with large value of time. 
Thus our models have good agreement with recent observations.     

%%%%%%%%%%%%%%%%%%%%%%%%%%%%%%%%%%%%%%%%%%%%%%%%%%%%%%%%%%%%%%%%%%%%%%%%%%%
%%%%%%%%%%%%%%%%%%%%%%%%%%   SECTION 5  %%%%%%%%%%%%%%%%%%%%%%%%%%%%%%
\section{Solution in the Polynomial Form}
Let $ L(S) = \frac{1}{2k_{3}\sqrt{S + k_{4}}}$, where $k_{3}$ and $k_{4}$ 
are  constants. \\ 

In this case, on integrating, Eq. (\ref{eq13}) gives the exact solution
\begin{equation}
\label{eq21}
S(t) = \alpha_{1}t^{2} + \alpha_{2}t + \alpha_{3},
\end{equation} 
where $\alpha_{1}$, $\alpha_{2}$ and $\alpha_{3}$ are arbitrary constants. \\

By using the same techniques as described in previous section $4$, we find 
the expressions for $\rho(t)$, $G(t)$ and $\Lambda(t)$:
\begin{equation}
\label{eq22}
\rho(t) = B (\alpha_{1}t^{2} + \alpha_{2}t + \alpha_{3})^{-3(1 + \gamma)},
\end{equation}  
\begin{equation}
\label{eq23}
G(t) = \frac{[2\alpha_{1}^{2} t^{2} + 2\alpha_{1}\alpha_{2} t  +\alpha_{2}^{2} + k - 
2\alpha_{1}\alpha_{3}](\alpha_{1}t^{2} + \alpha_{2}t + \alpha_{3})^{(1 + 3\gamma)}}
{4\pi(1 + \gamma)B},
\end{equation} 
\begin{equation}
\label{eq24} \Lambda = \frac{(1 + 3\gamma)\{(2\alpha_{1} t + \alpha_{2})^{2} + k \} 
+ 4\alpha_{1}(\alpha_{1}t^{2} + \alpha_{2}t + \alpha_{3})}{(1 + \gamma)
(\alpha_{1}t^{2} + \alpha_{2}t + \alpha_{3})^{2}},
\end{equation} 
where $B$ is an integrating constant.
%%%%%%%%%%%%%%%%%%% Figure 1 %%%%%%%%%%%%
\begin{figure}[htbp]
\centering
\includegraphics[width=8cm,height=8cm,angle=0]{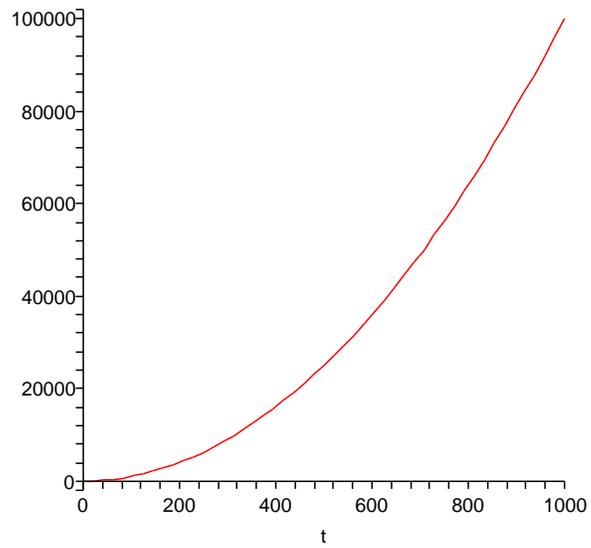}
\caption{The plot of scale factor $S(t)$ vs time with parameters 
$\alpha_{1} = 0.1$, $\alpha_{2} = 0.01$ and $\alpha_{3} = 1.0$}
\end{figure}
%%%%%%%%%%%%%%%%%%%%%%%%%%%%%%%%%% %%%%%%%%
%%%%%%%%%%%%%%%%%%% Figure 2 %%%%%%%%%%%%
\begin{figure}[htbp]
\centering
\includegraphics[width=8cm,height=8cm,angle=0]{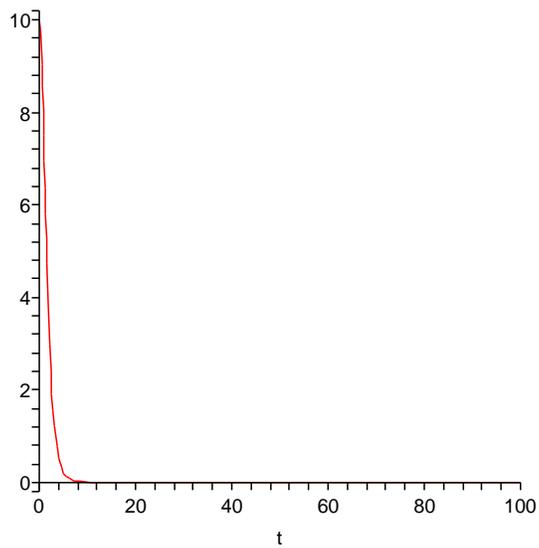}
\caption{The plot of energy density $\rho(t)$ vs time with parameters 
$\alpha_{1} = 0.1$, $\alpha_{2} = 0.01$, $\alpha_{3} = 1.0$, 
$\gamma = 0.0$ and $B = 1.0$}
\end{figure}
%%%%%%%%%%%%%%%%%%%%%%%%%%%%%%%%%% %%%%%%%%
%%%%%%%%%%%%%%%%%%% Figure 3 %%%%%%%%%%%%
\begin{figure}[htbp]
\centering
\includegraphics[width=8cm,height=8cm,angle=0]{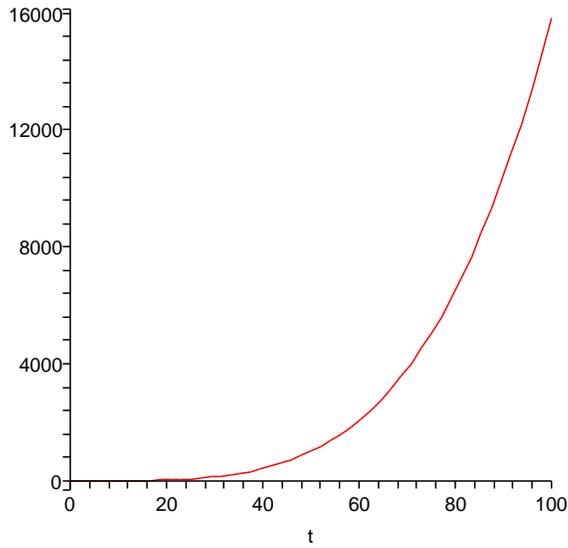}
\caption{The plot of gravitational term $G(t)$ vs time with parameters $k
= 1.0$, $\alpha_{1} = 0.1$, $\alpha_{2} = 0.01$, $\alpha_{3} = 1.0$, 
$\gamma = 0.0$ and $B = 1.0$}
\end{figure}
%%%%%%%%%%%%%%%%%%%%%%%%%%%%%%%%%% %%%%%%%%
%%%%%%%%%%%%%%%%%%% Figure 4 %%%%%%%%%%%%
\begin{figure}[htbp]
\centering
\includegraphics[width=8cm,height=8cm,angle=0]{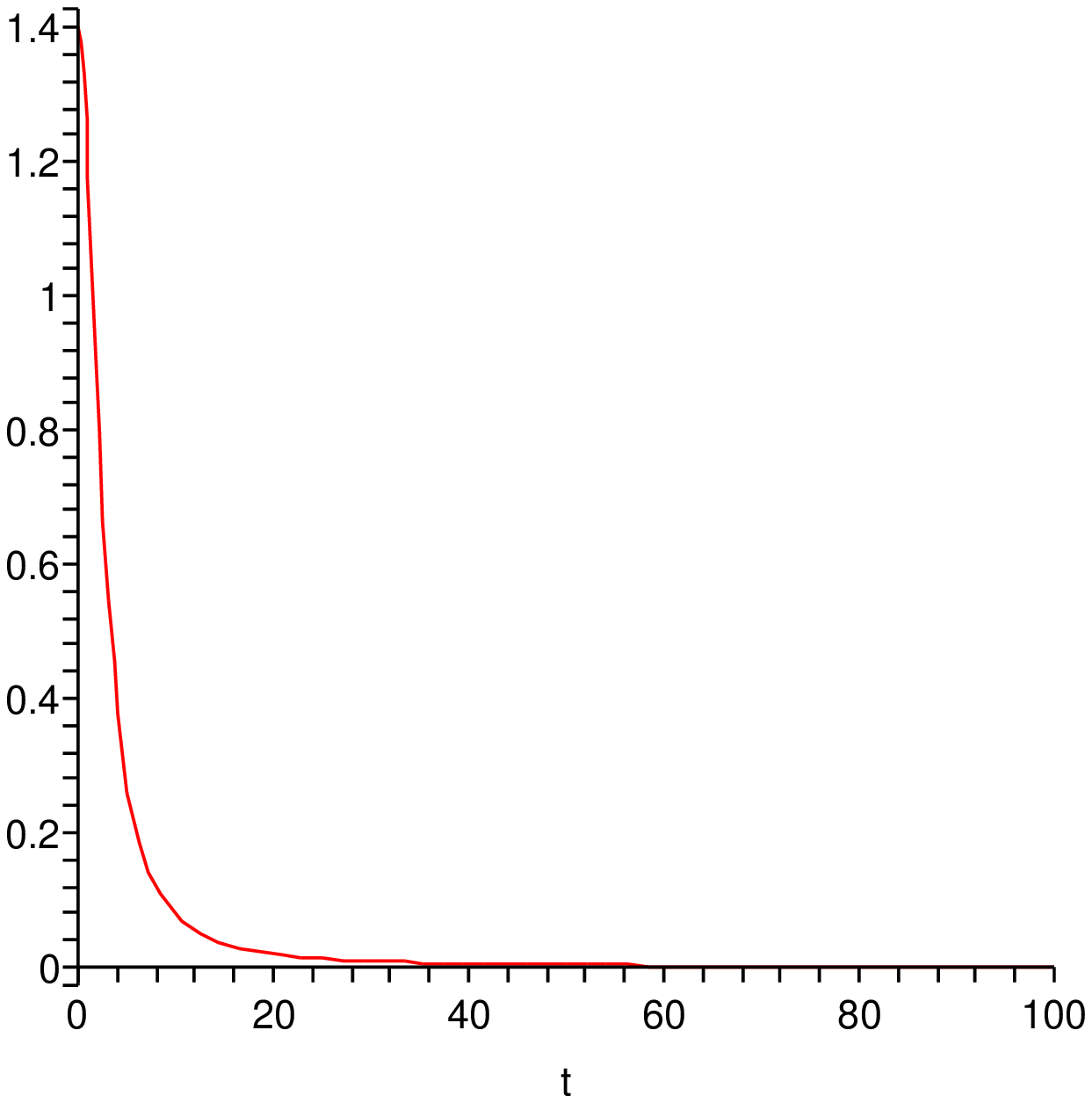}
\caption{The plot of cosmological term $\Lambda(t)$ vs time with parameter $k
= 1.0$, $\alpha_{1} = 0.1$, $\alpha_{2} = 0.01$, $\alpha_{3} = 1.0$ and 
$\gamma = 0.0$}
\end{figure}
%%%%%%%%%%%%%%%%%%%%%%%%%%%%%%%%%% %%%%%%%%
From Eq. (\ref{eq21}), since scale factor can not be negative, we find $S(t)$ is 
positive if $\alpha_{1}$, $\alpha_{2}$ and $\alpha_{3}$ are all greater the zero. 
From Figure $1$, it can be seen that in the early stages of the universe, i.e., 
near $t = 0$, the scale factor of the universe had been approximately constant and 
had increases very slowly. At a specific time the universe has exploded suddenly and 
it has expanded to large scale. This fits with big bang scenario. \\

From Eq. (\ref{eq22}), we observe that $\rho(t) > 0$ for $B > 0$. We also see that 
$\rho(t)$ is a decreasing function of time. Figure $2$ clearly shows this behaviour 
of $\rho(t)$. \\

From Eq. (\ref{eq23}), we note that $G(t)$ is an increasing function of time. Figure 
$3$ also shows this behaviour of $G$ as increasing function of $t$. From Eq. 
(\ref{eq26}), we observe that the cosmological term is a decreasing function of 
time and it approaches a small value as time progresses (i.e. the present epoch), 
which explains the small and positive value of $\Lambda$ at present 
(Perlmutter {\it et al.} \cite{ref16}; Riess {\it et al.} \cite{ref17,ref38}; 
Knop {\it et al.} \cite{ref21} Garnavich {\it et al.} \cite{ref19}; Schmidt 
{\it et al.} \cite{ref21}). Figure $4$ clearly shows this behaviour of $\Lambda$ 
as decreasing function of time. \\

The expressions for $\rho(t)$, $G(t)$ and $\Lambda(t)$ for dust universe 
($p = 0, \rho > 0$), Zel'dovich universe ($p = \rho$), and radiative universe 
($p = \frac{\rho} {3}$) can be obtained by putting $\gamma = 0, 1, \frac{1}{3}$ in 
Eqs. (\ref{eq22}), (\ref{eq23}) and (\ref{eq24}) respectively.  
%%%%%%%%%%%%%%%%%%%%%%%%%%%%%%%%%%%%%%%%%%%%%%%%%%%%%%%%%%%%%%%%%%%%%%%%%%%%%%
%%%%%%%%%%%%%%%%%%%  SECTION 6  %%%%%%%%%%%%%%%%%%%%%%%%%%%%%%%%%%%%%%%%%%

\section{Solution in the Sinusoidal Form}
Let $ L(R) = \frac{1}{\beta\sqrt{1 - R^{2}}}$, where $\beta$ is constant. \\

In this case, on integrating, Eq. (\ref{eq13}) gives the exact solution
\begin{equation}
\label{eq25}
S = M\sin(\beta t) + N\cos(\beta t) + \beta_{1},
\end{equation} 
where $M$, $N$ and $\beta_{1}$ are constants.

By using the same techniques as described in previous section $4$, we find the 
expressions 
for $\rho(t)$, $G(t)$ and $\Lambda(t)$:
\begin{equation}
\label{eq26}
\rho(t) = B [M\sin{\beta t} + N\cos{\beta t} + \beta_{1}]^{-3(1 + \gamma)},
\end{equation}  
\begin{equation}
\label{eq27}
G(t) = \frac{[\beta^{2}\{(M^{2} + N^{2}) - \beta_{1}(M\sin{\beta t} + 
N\cos{\beta t})\} + k][M\sin{\beta t} + N\cos{\beta t + \beta_{1}}]
^{(1 + 3\gamma)}}
{4\pi(1 + \gamma)B},
\end{equation} 
\[
\Lambda = \frac{(1 + 3\gamma)[\beta^{2}(M\cos{\beta t} - N\sin{\beta t})^{2} + k]}
{(1 + \gamma)[M\sin{\beta t} + N \cos{\beta t} + \beta_{1}]^{2}}
\]
\begin{equation}
\label{eq28} + \frac{2\beta^{2}(M\sin{\beta t} + N\cos{\beta t})}{(1 + \gamma)
[M\sin{\beta t} + N \cos{\beta t} + \beta_{1}].}  
\end{equation}
%%%%%%%%%%%%%%%%%%% Figure 5 %%%%%%%%%%%%
\begin{figure}[htbp]
\centering
\includegraphics[width=8cm,height=8cm,angle=-90]{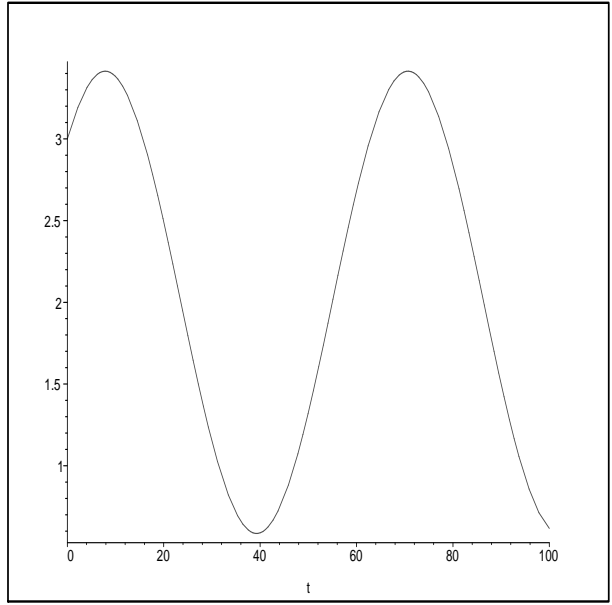}
\caption{The plot of scale factor $S(t)$ vs time with parameter 
$M = 1.0$, $N = 1.0$, $\gamma = 0.0$, $\beta = 0.10$ 
and $\beta_{1} = 2.0$}
\end{figure}
%%%%%%%%%%%%%%%%%%%%%%%%%%%%%%%%%% %%%%%%%%
%%%%%%%%%%%%%%%%%%% Figure 6 %%%%%%%%%%%%
\begin{figure}[htbp]
\centering
\includegraphics[width=8cm,height=8cm,angle=-90]{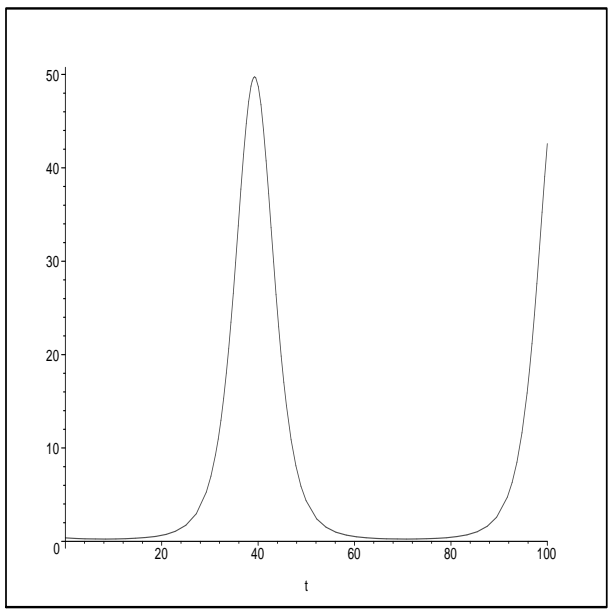}
\caption{The plot of energy density $\rho(t)$ vs time with parameter 
$M = 1.0$, $N = 1.0$, $B = 10.0$, $\gamma = 0.0$, $\beta = 0.10$ 
and $\beta_{1} = 2.0$} 
\end{figure}
%%%%%%%%%%%%%%%%%%%%%%%%%%%%%%%%%% %%%%%%%%
%%%%%%%%%%%%%%%%%%% Figure 7 %%%%%%%%%%%%
\begin{figure}[htbp]
\centering
\includegraphics[width=8cm,height=8cm,angle=-90]{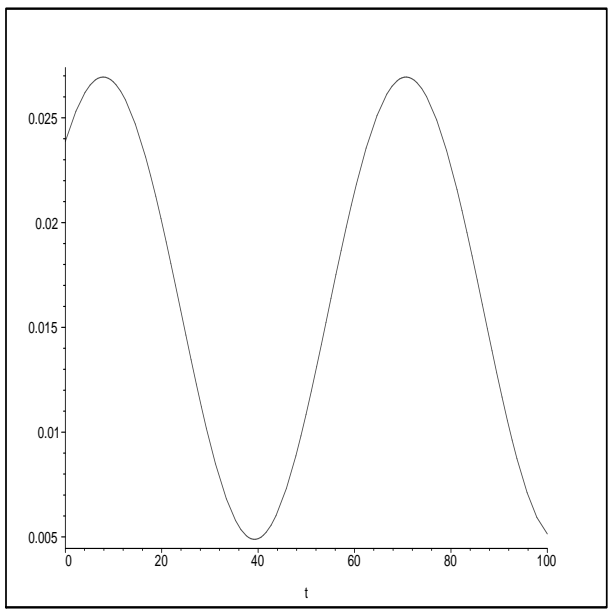}
\caption{The plot of gravitational term $G(t)$ vs time with parameter $k
= 1.0$, $M = 1.0$, $N = 1.0$, $B = 10.0$, $\gamma = 0.0$, $\beta = 0.10$ 
and $\beta_{1} = 2.0$}  
\end{figure}
%%%%%%%%%%%%%%%%%%%%%%%%%%%%%%%%%% %%%%%%%%
%%%%%%%%%%%%%%%%%%% Figure 8 %%%%%%%%%%%%
\begin{figure}[htbp]
\centering
\includegraphics[width=8cm,height=8cm,angle=-90]{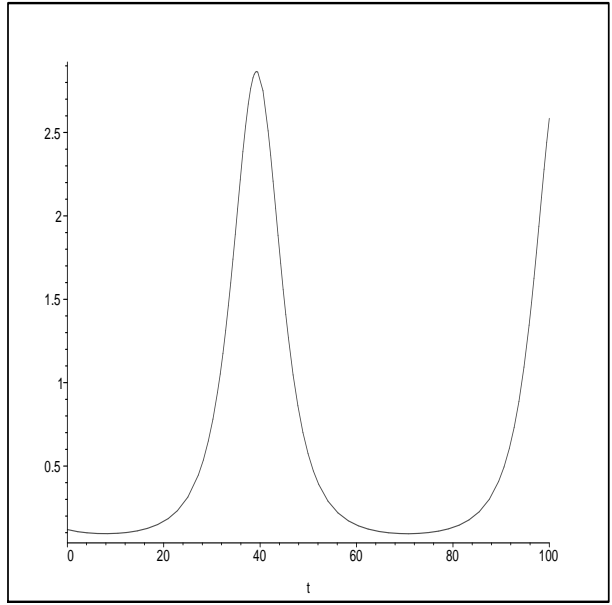}
\caption{The plot of cosmological term $\Lambda(t)$ vs time with parameter $k
= 1.0$, $M = 1.0$, $N = 1.0$, $B = 10.0$, $\gamma = 0.0$, $\beta = 0.10$ 
and $\beta_{1} = 2.0$}  
\end{figure}
Since, in this case, we have many alternatives for choosing values of $M$, $N$, 
$\beta$, $\beta_{1}$, it is just enough to look for suitable values of these 
parameters, such that the physical initial and boundary conditions are satisfied. 
We are trying to find feasible interpretation and situations relevant to this case. \\

In Figure $5$ we plot scale factor versus time. It is observed here that at early 
stages of the universe, the scale of the universe increases gently and then 
decreases sharply, and afterwards it will oscillate for ever. \\

In Figure $6$ we plot energy density $\rho(t)$ versus time and in Figure $8$ we note 
the run of cosmological constant $\Lambda(t)$ versus time. From these figures we 
conclude that at the early stages of the universe the matter is created as a result 
of loss of vacuum energy and at a particular epoch it has started to oscillate for 
ever due to sinusoidal property. In Figure $7$ we plot the $G(t)$ versus time. From 
it we conclude that in the beginning at the early stages of the universe $G(t)$ 
increases gently with time and at a particular epoch it decreases sharply, and 
afterwards it will oscillate for ever. It is worth to mention here that in this 
case oscillation takes place in positive quadrant which is physically relevant. \\

The expressions for $\rho(t)$, $G(t)$ and $\Lambda(t)$ for dust universe 
($p = 0, \rho > 0$), Zel'dovich universe ($p = \rho$), and radiative universe 
($p = \frac{\rho}{ 3}$) can be obtained by putting $\gamma = 0, 1, \frac{1}{3}$ in 
Eqs. (\ref{eq26}), (\ref{eq27}) and (\ref{eq28}) respectively.    
%%%%%%%%%%%%%%%%%%%%%%%%%%%%%%%%%% %%%%%%%% 
\section{Discussion and Conclusion}
In this paper we have described a new class of FRW cosmological models of the universe 
with a perfect fluid as the source of matter by applying a variable deceleration 
parameter. The ``constants'' $G$ and $\Lambda$ are allowed to depend on the 
cosmic time. Generally, the models are expanding, non-shearing and isotropic 
in nature. \\

The cosmological terms in the models in Sections $4$, $5$ are a decreasing function 
of time and this approaches a small value as time increases (i.e. present epoch). 
The value of the cosmological ``terms'' for these models are found to be small and 
positive which is supported by the results from recent supernovae Ia observations  
(Perlmutter {\it et al.} \cite{ref16}; Riess {\it et al.} \cite{ref17,ref38}; 
Knop {\it et al.} \cite{ref37};  Garnavich {\it et al.} \cite{ref19}; Schmidt {\it et al.} 
\cite{ref21}). The cosmological term in Section $6$ also decreases while time increases 
to a specific instant. During this period as we can understand from Figure $8$, 
we will have enough matter creation to force the universe to oscillate for ever due 
to sinusoidal property of $\Lambda$. This means we always have annihilation and 
creation of matter permanently. At this point one more sentence may be added to 
our discussion and i.e. as the graphs for $\Lambda$ and $\rho$ in this case the 
explosion of the universe at the early stages of its creation has been only a 
consequence of matter creation. Thus, the implications of time varying $\Lambda$ 
and $G$ are important to study the early evolution of the universe. \\

    \section*{Acknowledgements}
One of the authors (A. Pradhan) thanks the Inter-University Centre for Astronomy 
and Astrophysics (IUCAA), Pune,  India for providing  facility under associateship 
programmes where a part of this work was carried out. 
%\newline
%\nonumsection{References}

\end{document}